\documentstyle[psfig,12pt,twoside,fleqn,espcrc1]{article}

\newcommand{\AmS}{{\protect\the\textfont2
  A\kern-.1667em\lower.5ex\hbox{M}\kern-.125emS}}

\begin{document}
\title{\vspace*{-2cm}
\hfill MKPH-T-97-32\\
{\bf On the Gerasimov-Drell-Hearn sum rule for the deuteron}
\thanks{Supported by the Deutsche Forschungsgemeinschaft (SFB 201)}
\thanks{Contribution to Few-Body XV, Groningen (1997)}
} 

\author{H.\ Arenh\"ovel, G.\ Kre\ss, R.\ Schmidt and P.\ Wilhelm
\address{
Institut f\"{u}r Kernphysik, Johannes Gutenberg-Universit\"{a}t,
       D-55099 Mainz, Germany}}
\maketitle

\begin{abstract} 
The Gerasimov-Drell-Hearn sum rule is evaluated for the deuteron by
explicit integration up to 550 MeV including contributions 
from the photodisintegration channel and from coherent and incoherent 
single pion production as well. The photodisintegration channel converges 
fast enough in this energy range and gives a large negative contribution, 
essentially from the $^1S_0$ resonant state near threshold. 
Its absolute value is about the same size as the 
sum of proton and neutron GDH values. It is only partially cancelled by the 
single pion production contribution. But the incoherent channel 
has not reached convergence at 550 MeV. 
\end{abstract}

\section{INTRODUCTION}
\label{sec1}

The Gerasimov-Drell-Hearn (GDH) sum rule connects the anomalous magnetic moment 
of a particle with the energy weighted integral - henceforth denoted by 
$I^{GDH}$ - from threshold up to infinity over the difference of the total 
photoabsorption cross sections for circularly polarized photons on a target 
with spin parallel and antiparallel to the spin of the photon. 
In detail it reads for a particle of mass $m$, charge $eQ$, anomalous 
magnetic moment $\kappa$ and spin $S$ 
\begin{equation}
I^{GDH}=4\pi^2\kappa^2\frac{e^2}{m^2}\,S
=\int_0^\infty \frac{dk}{k}
\left(\sigma ^P(k)-\sigma ^A(k)\right)
\,,\label{gdh}
\end{equation}
where $\sigma ^{P/A}(k)$ denote the total absorption cross sections for
circularly polarized photons on a target with spin parallel and antiparallel 
to the photon spin, respectively. The anomalous magnetic moment is 
defined by the total magnetic moment operator of the particle 
$\vec M = (Q+\kappa)(e/m)\vec S$.
This sum rule gives a very interesting relation between a 
ground state property of a particle and an integral property of 
its whole excitation spectrum showing that the existence of a 
nonvanishing anomalous magnetic moment points directly to an 
internal structure. 

The GDH sum rule has first been derived by Gerasimov \cite{Ger65} and, 
shortly afterwards, independently by Drell and Hearn \cite{DrH66}. 
It is based on the low energy theorem of Low \cite{Low54} and Gell-Mann 
and Goldberger \cite{GeG54} for a spin one-half particle which later has 
been generalized to arbitrary spin \cite{LaC60,Sai69,Fri77}, and on 
the assumption of an unsubtracted dispersion 
relation for the difference of the elastic forward scattering
amplitudes for circularly polarized photons and a completely polarized
target with spin parallel and antiparallel to the photon spin. 

\begin{figure}[h]
\centerline{\psfig{figure=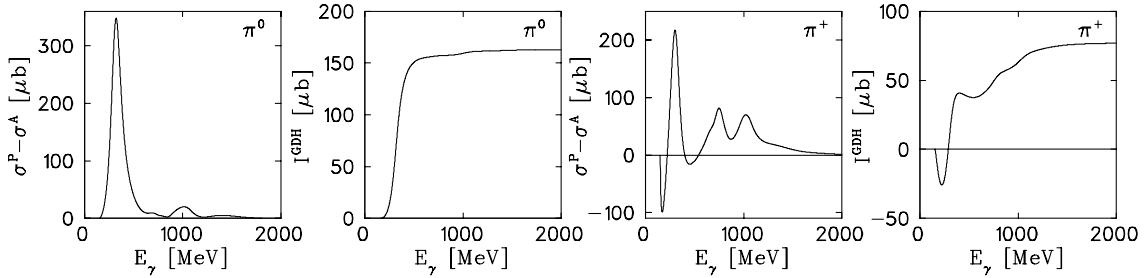,width=15cm,angle=0}}
\vspace*{-1cm}
\caption{
Left two panels: spin asymmetry and \protect{$I^{GDH}_p$} for $\pi^0$ 
photoproduction on the proton; Right two panels: same for $\pi^+$ 
photoproduction (from fit SM95 of \protect{\cite{VPI}}). }
\label{proc_fig1}
\vspace*{-.8cm}
\end{figure}

Since proton and neutron have large anomalous magnetic moments, one finds 
large GDH sume rule predictions for them, i.e., $I^{GDH}_p=204.8\,\mu$b 
for the proton and $I^{GDH}_n=233.2\,\mu$b for the neutron. 
Unfortunately, this sum rule has never been 
verified by a direct measurement. Early evaluation by Karliner \cite{Kar73}, 
based on a multipole analysis of experimental data on pion 
photoproduction on the nucleon, did not give 
conclusive results, and even 
present day data do not allow a definite answer as to its validity (see e.g. 
\cite{SaW94}). 
We show in Fig.\ \ref{proc_fig1} the 
contributions to the spin asymmetry for the proton and the corresponding $
I^{GDH}_p$ integrated up to 2 GeV. One finds $I^{GDH}_p(2\,\mbox{GeV})= 239
\,\mu$b and $I^{GDH}_n(2\,\mbox{GeV})= 168\,\mu$b for the VPI fit SM95 
\cite{VPI} which deviate significantly from the above sum rule values. 
\vspace*{-.7cm}

For the deuteron, one observes a very interesting feature. Since it has 
isospin zero, one finds a very small anomalous magnetic moment 
$\kappa_d=-.143$ resulting in a GDH prediction of $I^{GDH}_d = 0.65\,\mu$b, 
which is more than two orders of magnitude smaller 
than the nucleon values. On the other hand, the incoherent pion production on 
the deuteron is dominated by the quasifree production resulting in a 
contribution to the GDH value of roughly the sum of the proton and 
neutron GDH values, i.e., 438 $\mu$b. Additional contributions arise from 
the coherent $\pi^0$ production channel. Therefore, 
one needs a large negative contribution of 
about the same size for cancellation. Indeed, from the photodisintegration 
channel, not present for the nucleon, one finds at very low energies near 
threshold a sizeable negative contribution from the $M1$-transition to the 
resonant $^1S_0$ state, because this state can only be reached if the spins 
of photon and deuteron are antiparallel \cite{BaD67}. 

Recently, we have evaluated the GDH sum rule for the deuteron by explicit 
integration up to a photon energy of 550 MeV including the 
photodisintegration channel as well as coherent and incoherent single 
pion photoproduction channels \cite{ArK97}. The 
reason for the cut-off of the upper integration limit is that around this 
energy the two-pion contribution starts to become significant \cite{SaW94} 
for which we do not have a reliable model.

\section{THE GDH SUM RULE FOR THE DEUTERON}
\label{sec3}
For the deuteron, three contributions have been included: (i) the 
photodisintegration channel $\gamma d \rightarrow n p$, (ii) the coherent 
pion production $\gamma d \rightarrow \pi^0 d$, and (iii) the 
incoherent pion production $\gamma d \rightarrow \pi N N$. 
We will now discuss the three contributions separately. 

\subsection{Photodisintegration}
\label{sec3a}
Near the break-up threshold, only E1 and M1 contribute significantly. 
However, E1 is largely cancelled in the spin asymmetry. Thus, at 
low energies only $M1$ transitions remain, essentially to $^1S_0$ and 
$^3S_1$ states. Of these, the $^1S_0$ contributions is dominant because 
of the large isovector part of the $M1$-operator coming from the large 
isovector anomalous magnetic moment of the nucleon. It is particularly 
strong close to break-up threshold where the 
$^1S_0$ state is resonant. It can only be reached by the antiparallel spin 
combination resulting in a strong negative contribution to the GDH sum rule. 

The photodisintegration channel is evaluated within 
the nonrelativistic framework as is described in detail in
Ref.~\cite{ArS91} but with inclusion of the most important relativistic 
contributions. All electric and magnetic multipoles up to 
the order $L=4$ are considered. The calculation is based on 
the realistic Bonn potential (r-space version) \cite{MaH87}. 
In the current operator we distinguish the one-body currents with Siegert 
operators (N), explicit meson exchange contributions (MEC) beyond the
Siegert operators, essentially from $\pi$- and $\rho$-exchange, 
contributions from isobar configurations of the wave functions (IC), 
calculated in the impulse approximation \cite{WeA78}, and leading order 
relativistic contributions (RC) of which the spin-orbit current is 
by far the most dominant part. 
\vspace*{-.7cm}

\begin{figure}[h]
\centerline{\psfig{figure=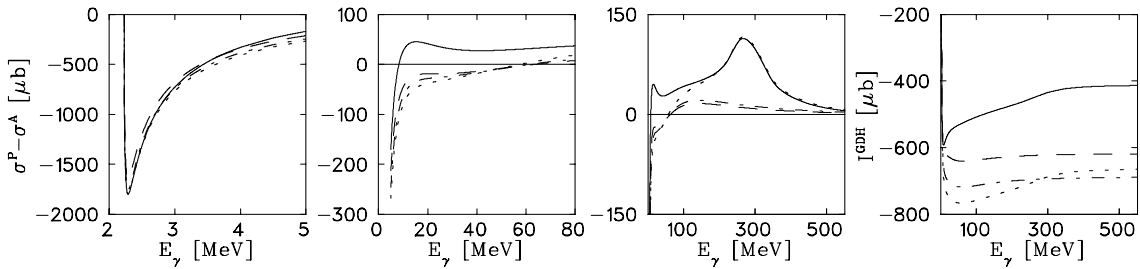,width=15cm,angle=0}}
\vspace*{-1cm}
\caption{
Contribution of deuteron photodisintegration to the GDH sum rule. 
Three  left panels: difference of the cross sections 
in various energy regions; right panel: 
$I^{GDH}_{\gamma d \to np}$ as function of the upper integration energy. Dashed 
curves: N, dash-dot: N+MEC, dotted: N+MEC+IC, and full curves N+MEC+IC+RC.}
\label{proc_fig2}
\vspace*{-.8cm}
\end{figure}

\vspace*{-.7cm}

\begin{table}[h]
\newlength{\digitwidth} \settowidth{\digitwidth}{\rm 0}
\catcode`?=\active \def?{\kern\digitwidth}
\caption{Contributions of deuteron photodisintegration to the 
GDH integral up to 550 MeV in $\mu$b.}
\label{tabdis}
\begin{tabular*}{\textwidth}{@{}c@{\extracolsep{\fill}}cccc}
\hline
  N & N+MEC & N+MEC+IC & N+MEC+IC+RC\\ 
   \hline
 $-619$ & $-689$ & $-665$ & $-413$ \\
\hline
\end{tabular*}
\vspace*{-.9cm}
\end{table}

The results are summarized in Fig.\ \ref{proc_fig2}, where the spin 
asymmetry and the GDH integral are shown. The GDH values are listed in 
Tab.\ \ref{tabdis}. One readily notes the huge negative contribution from 
the $^1S_0$-state at low energies (see the left panel of Fig.\ 
\ref{proc_fig2}). Here, the effects from MEC 
are relatively strong resulting in an enhancement of the negative value by 
about 15 percent. Isobar effects are significant in the region of 
the $\Delta$-resonance, as expected. They give a positive contribution, but 
considerably smaller in absolute size than MEC. The largest positive 
contribution stems from RC in the energy region up to about 100 MeV 
(see the second left panel of Fig.\ \ref{proc_fig2})
reducing the GDH value in absolute size by more than 30 percent. This 
strong influence from relativistic effects is not surprising in view of the 
fact, that the correct form of 
the low energy expansion of the forward Compton scattering amplitude 
is only obtained if leading order relativistic contributions 
are included \cite{Fri77}. 
The total sum rule value from the photodisintegration channel is 
$I^{GDH}_{\gamma d \to np}(550\,\mbox{MeV})=-413\,\mu$b. Its absolute value 
almost equals within less than ten percent the sum of the free proton 
and neutron values. This may not be accidental since the large value is 
directly linked to the nucleon anomalous magnetic moment as is demonstrated 
by the fact that one finds indeed a very small but positive value 
$I^{GDH}_{\gamma d \to np}(550\,\mbox{MeV})=7.3\,\mu$b if the nucleon 
anomalous magnetic moment is switched off in the e.m.\ one-body current 
operator. 

\subsection{Coherent pion production}
\label{sec3b}

The theoretical model used to calculate the contribution of the
coherent pion production channel is described in detail in Ref.\ 
\cite{WiA95}. The reaction is dominated by the
magnetic dipole excitation of the $\Delta$ resonance resulting in 
a strong positive $I^{GDH}_{\gamma d \to d\pi^0}$ contribution, because the 
$\Delta$-excitation is favoured if photon and nucleon spins are parallel 
compared to the antiparallel situation. The model takes into
account pion rescattering by solving a system of coupled equations for
the N$\Delta$, NN$\pi$ and NN channels.  The most important
rescattering mechanism is due the successive excitation and decay of
the $\Delta$ resonance.  The inclusion of the rescattering effects is 
important and leads in general to a significant reduction of the cross 
section in reasonable agreement with the differential cross 
section data available in the $\Delta$ region. We find a sizeable 
positive contribution from the $\Delta$-excitation giving a value 
$I^{GDH}_{\gamma d \to d\pi^0}(550\,\mbox{MeV})=63\,\mu$b, and 
satisfactory convergence is achieved.

\subsection{Incoherent pion production}
\label{sec3c}
The calculation of the $\gamma d \rightarrow \pi NN$ contributions to
the GDH integral is based on the spectator nucleon approach discussed 
in \cite{ScA96} and details may be found there. 
In this framework, the reaction proceeds via the pion production on one 
nucleon while the other nucleon acts merely as a spectator. 
Although the spectator model 
does not include any final state interaction except for the resonant 
$M_{1+}^{3/2}$ multipole, it gives quite a good description of available 
data on the deuteron. 
Due to the neglect of $NN$ rescattering in the final 
state, there is some double counting with respect to the coherent process. 
\vspace*{-.5cm}

\begin{figure}[h]
\centerline{\psfig{figure=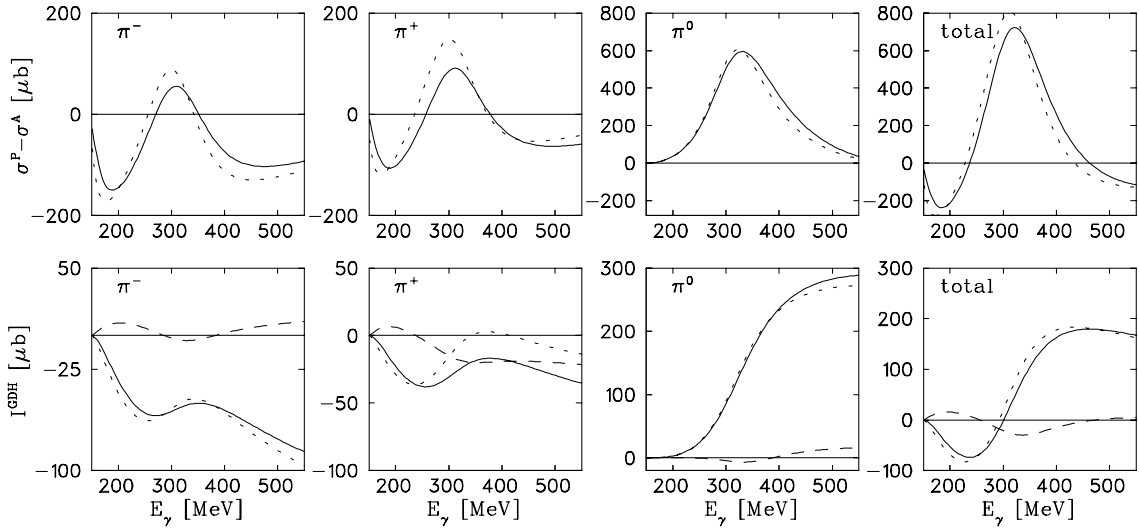,width=15cm,angle=0}}
\vspace*{-1cm}
\caption{
Contribution of the incoherent $\pi$ production to the GDH sum rule for the 
deuteron and the nucleon. Upper part: difference of the cross sections; 
lower part: $I^{GDH}_{\gamma d \to NN\pi}$ as function of the upper integration energy. Full 
curves for the deuteron, dotted curves for the nucleon. In the case of $\pi^0$ 
production, the dotted curve shows the summed proton and neutron 
contributions. The dashed curves show the appropriate differences 
$I^{GDH}_{\gamma d \to NN\pi}-I^{GDH}_p-I^{GDH}_n$.} 
\label{proc_fig3}
\vspace*{-.8cm}
\end{figure}

The results are collected in Fig.\ \ref{proc_fig3}. The upper part shows the 
individual contributions from the different charge states of the pion 
and their total sum to 
the cross section difference for pion photoproduction on both the deuteron 
and for comparison on the nucleon. One notes qualitatively a similar 
behaviour although the maxima and minima are smaller and also slightly 
shifted towards higher energies for the deuteron. In the lower part of 
Fig.\ \ref{proc_fig3} the corresponding GDH integrals are shown. A large 
positive 
contribution comes from $\pi^0$-production whereas the charged pions give a 
negative but - in absolute size - smaller contribution to the GDH value. Up to 
an energy of 550 MeV one finds for the total contribution of the incoherent 
pion production channels a value 
$I^{GDH}_{\gamma d \to NN\pi}(550\,\mbox{MeV})=167\,\mu$b which is remarkably 
close to the sum of the neutron and proton values for the given elementary 
model $I^{GDH}_n(550\,\mbox{MeV})+I^{GDH}_p(550\,\mbox{MeV})=163\,\mu$b. 
This fact is also indicated by the dashed curves in the lower part of 
Fig.\ \ref{proc_fig3} which represent the appropriate differences of 
$I^{GDH}_{\gamma d \to NN\pi}-I^{GDH}_p-I^{GDH}_n$. 
It underlines again that the total cross 
section is dominated by the quasifree process. However, as is 
evident from Fig.\ \ref{proc_fig3}, convergence is certainly not reached at this 
energy. Furthermore, the elementary pion production operator had been 
constructed primarily to give a realistic description of the $\Delta$ 
reonance region. In fact, it underestimates the GDH inegral up to 550 MeV 
by about a factor two compared to a corresponding evaluation based on a 
multipole analysis of experimental pion photoproduction data. 
But the important result is, that the total GDH contribution from the 
incoherent process is very close to the sum of the free proton and neutron GDH 
integrals which will remain valid for an improved elementary production 
operator. 

\section{SUMMARY AND CONCLUSIONS}
\label{sec4}
The contributions from all three channels and their sum are listed in Tab.\ 
\ref{tab1}. A very interesting and important result is the large negative 
contribution from the photodisintegration channel near and not too far 
above the break-up threshold with surprisingly large relativistic effects 
below 100 MeV. Hopefully, this low energy feature of 
the GDH sum rule could be checked experimentally in the near future. 
\vspace*{-.6cm}

\begin{table}[h]
\catcode`?=\active \def?{\kern\digitwidth}
\label{tab1}
\caption{
GDH contributions for the deuteron integrated up to 550 MeV in $\mu$b.}
\begin{tabular*}{\textwidth}{@{}c@{\extracolsep{\fill}}ccccc}
\hline
   $\gamma d \to np$ & $\gamma  d \to d \pi^0$ & $\gamma d \to np\pi^0$ 
  &$\gamma d \to nn\pi^+$   & $\gamma d \to pp\pi^-$   &total\\
\hline
  $-413$ &   63 &  288 &  $-35$ &  $-86$ & $-183$ \\
\hline
\end{tabular*}
\vspace*{-.8cm}
\end{table}

For the total GDH value from explicit integration up to 550 MeV, we find a 
negative value $I^{GDH}_d(550\,\mbox{MeV})=-183\,\mu$b. However, as we have 
mentioned above, some uncertainty lies in the contribution of the incoherent
pion production channel because of shortcomings of the model of the
elementary production amplitude above the $\Delta$ resonance. If we use 
instead of the model value $I^{GDH}_{\gamma d \to NN\pi}(550\,\mbox{MeV})=
167\,\mu$b 
the sum of the GDH values of neutron and proton by integrating the 
cross section difference obtained from a multipole analysis of experimental 
data (fit SM95 from \cite{VPI}), giving $I^{GDH}_n(550\,\mbox{MeV})
+I^{GDH}_p(550\,\mbox{MeV})=331\,\mu$b, we find for the deuteron 
$I^{GDH}_d(550\,\mbox{MeV})=-19\,\mu$b, which we consider a more realistic 
estimate. Since this value is still negative, a positive contribution of 
about the same size should come from contributions at higher energies in 
order to fulfil the small GDH sum rule for the deuteron, provided that the 
sum rule is valid. These contributions should come from the incoherent 
single pion production above 550 MeV because here convergence had not been 
reached in contrast to the other two channels, photodisintegration and 
coherent pion production, and in addition, from multipion production. 

It remains as a task for future research to improve the elementary pion 
photoproduction operator above the $\Delta$ resonance. But for this 
also precise data on $\sigma^P-\sigma^A$ from a direct measurement is 
urgently needed. 
Furthermore, for the reaction on the deuteron, the influence of final state 
interaction has to be investigated, too. Because the large cancellation 
between the various contributions requires quite a high degree of precision 
for the theoretical description. For this reason, also at least two-pion 
production contributions have to be considered in order to obtain more 
reliable predictions at higher energies. 

Finally, we would like to add that even at this incomplete stage due 
to the cut-off of the upper integration limit at 550 MeV, the results are
sufficiently interesting. This is based on the fact that not
only the sum rule value is of interest, but that also the asymmetry
$\sigma_P-\sigma_A$ itself as function of the photon energy is a very
important quantity which deserves detailed investigations for the various 
channels. The reason for 
this is that this asymmetry contains very interesting physics with respect 
to the hadronic structur of the system describing its optical activity which 
reflects an internal screw-like or chiral structure.

\end{document}